\documentclass[10pt,journal,compsoc]{IEEEtran}
\usepackage{graphicx}
\usepackage{mathrsfs}
\usepackage{amsfonts,amssymb}
\usepackage{bm}
\usepackage{color}
\usepackage{graphicx,amssymb,booktabs}
\usepackage{graphicx}
\usepackage{bm}
\usepackage{array}
\usepackage{amsmath}
\usepackage{float}

\usepackage{cite}

\ifCLASSINFOpdf
\else
\fi

\begin{document}
%
\title{D-UNet: a dimension-fusion U shape network for chronic stroke lesion segmentation}
%
%
%
%

\author{Yongjin~Zhou,~\IEEEmembership{Member,~IEEE,}
        Weijian~Huang, 
        Pei~Dong, 
        Yong~Xia, 
        and~Shanshan~Wang,~\IEEEmembership{Member,~IEEE,}
\IEEEcompsocitemizethanks{\IEEEcompsocthanksitem This work was supported by funding from the National Natural Science Foundation of China (61601450, 61871371, and 81830056), Science and Technology Planning Project of Guangdong Province (2017B020227012) (Corresponding authors: Shanshan Wang)
\IEEEcompsocthanksitem Y. Z and W. H are with the School of Shenzhen University, Shenzhen
518060, China. E-mail: yjzhou@szu.edu.cn, 2170249218@email.szu.edu.cn
\IEEEcompsocthanksitem P. D is with School of Information Technologies, University of Sydney, NSW 2006, Australia.
\IEEEcompsocthanksitem Y. X is with School of Computer Science, Northwestern Polytechnical University, Xi’an 710072, China. 
\IEEEcompsocthanksitem S. Wang is with the Paul C. Lauterbur Research Center for Biomedical Imaging, Shenzhen Institutes of Advanced Technology, Chinese Academy of Sciences, Shenzhen 518055, China. E-mail: ss.wang@siat.ac.cn, sophiasswang@hotmail.com.
\IEEEcompsocthanksitem Code will be available at: https://github.com/SZUHvern/D-UNet/tree/master.}}

\IEEEtitleabstractindextext{%
\begin{abstract}
Assessing the location and extent of lesions caused by chronic stroke is critical for medical diagnosis, surgical planning, and prognosis. In recent years, with the rapid development of 2D and 3D convolutional neural networks (CNN), the encoder-decoder structure has shown great potential in the field of medical image segmentation. However, the 2D CNN ignores the 3D information of medical images, while the 3D CNN suffers from high computational resource demands. This paper proposes a new architecture called dimension-fusion-UNet (D-UNet), which combines 2D and 3D convolution innovatively in the encoding stage. The proposed architecture achieves a better segmentation performance than 2D networks, while requiring significantly less computation time in comparison to 3D networks. Furthermore, to alleviate the data imbalance issue between positive and negative samples for the network training, we propose a new loss function called Enhance Mixing Loss (EML). This function adds a weighted focal coefficient and combines two traditional loss functions. The proposed method has been tested on the ATLAS dataset and compared to three state-of-the-art methods. The results demonstrate that the proposed method achieves the best quality performance in terms of DSC = 0.5349\(\pm\)0.2763 and precision = 0.6331\(\pm\)0.295).
\end{abstract}

}

\maketitle

\IEEEdisplaynontitleabstractindextext

%
\IEEEpeerreviewmaketitle

\IEEEraisesectionheading{\section{Introduction}\label{sec:introduction}}

%
%
%
%
\IEEEPARstart{S}{troke} is the most common cerebrovascular disease and is one of the most common causes of death and disability worldwide\cite{1,2}. It is a group of diseases caused by a sudden cerebrovascular rupture or cerebrovascular infraction. The typical symptom of this disease is a focal neurological deficit, such as sudden seizures, language disorders, hemianopia, loss of feeling, etc.\cite{3}. These symptoms may develop into chronic diseases (such as dementia, hemiplegia, etc.), which can seriously affect the life quality of patients; these diseases consume a large part of social health care costs \cite{4}. At the subacute/chronic stages, effective rehabilitation can promote a long-term functional recovery. However, there have been few advances in large-scale neuroimaging-based stroke predictions at the subacute and chronic stages. The most common research scan is a high-resolution T1-weighted structural MRI. Researches using these types of images at the subacture/chronic stages have revealed promising biomarkers. These could potentially provide additional information, beyond behavioral assessments, to predict an individual’s likelihood of recovery for specific functions (e.g., motor, speech) and response to treatments \cite{5,6}. Thus far, measures that include the size, location, and overlap of the lesion with existing brain regions or structures, such as the corticospinal tract, have been successfully used as predictors of long-term stroke recovery and rehabilitation \cite{7}. However, a key barrier to correctly analyzing these large-scale stroke neuroimaging datasets to predict outcomes is the accurate segmentation of lesions. As manually-based annotations may no longer be suitable for a wide range of data requirements, there is a need for automatic segmentation tools for their analyses.
\begin{figure}[htbp]
	\centering
	\includegraphics[width=\linewidth,scale=1]{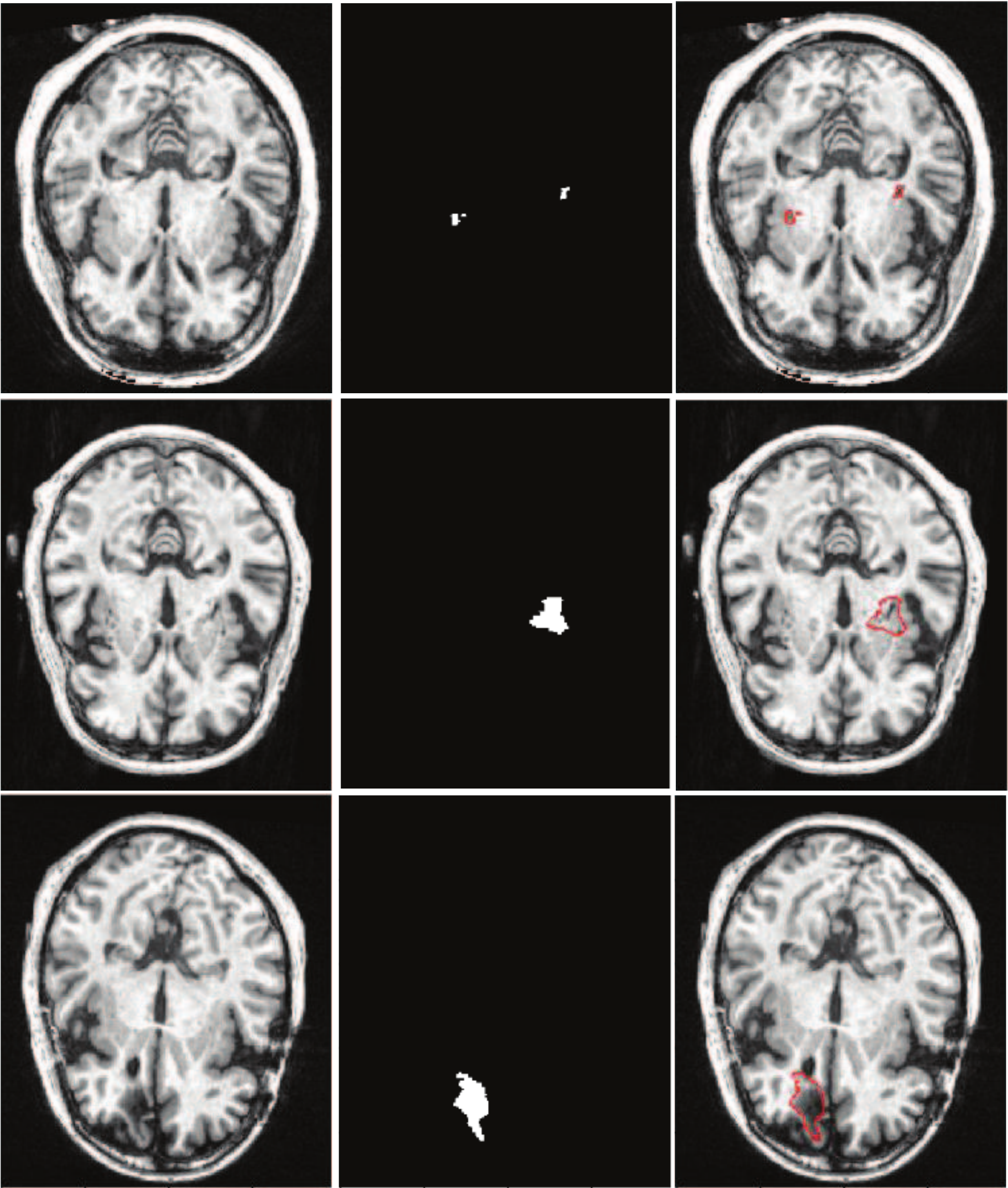}
	\caption{The MRI T1 sequence stroke image from the ATLAS dataset. The first column is the raw data, the second column is the gold standard from the hand-marked lesions by the doctor, and the third column is the combination of the first two columns. Strokes occur in different locations, with large differences in shape and unclear boundaries.} \label{fig1}
\end{figure}

Strokes occur in different locations, with large differences in shape and unclear boundaries as shown in Fig. 1. A public dataset, Anatomical Tracings of Lesions-After-Stroke (ATLAS), is  utilized  to illustrate this variability\cite{7}. Firstly, the segmentation performance is reduced by motion artifacts in the MRI images. Secondly, the position and shape of the lesions are significantly different owing to the existence of multiple subtypes of strokes. The lesion volume can vary from hundreds to tens of thousands of cubic millimeters depending on the severity of the disease, and the lesion area can occur in the cerebrum, cerebellum, and other areas of the brain. Finally, the boundaries of some lesions are not clear, and different clinicians may inconsistently label different lesion areas. Therefore, the accurate automated segmentation is a challenging problem.

To tackle these difficulties, researchers have made many efforts, including intensity threshold processing, region growth, and deformable models. However, these methods rely on the hand-crafted feature extraction by experts; they have a limited feature representation and low generalization performance. In recent years, with the rapid development of deep learning, convolutional neural networks (CNN) have proven to have great potential in the field of medical image analysis \cite{7,8,9,10,11,12,13,14,15,16,17}. The study of CNN is mainly based on two-dimensional (2D) and three-dimensional (3D) approaches: (1) In the 2D CNN approaches, the MRI volume data are converted into several planar slices and independently predict the lesion area of each slice. These ignore the spatial characteristics of the MRI data such that the predictions are discontinuous. (2) In the 3D CNN, approaches, spatial information is extracted for inference. However, due to their computational and storage requirements, the 3D CNN have been largely avoided.

In order to solve the problem of accurately automating the image segmentation, we propose a novel network called the Dimension-fusion-UNet (D-UNet). In this new model, the 3D spatial information in the MRI data is effectively utilized under the 2D framework of the subject and has low computing resource requirements. Our D-UNet has the following two technical achievements:

\textbf{Dimension fusion network:} First, in order to extract the information of consecutive slices from MRI data, we designed a novel downsampling block based on a UNet improvement. This improvement performs 3D and 2D feature extraction on a small number of consecutive slices in the early stage of the network. Then, in a novel way, their respective feature maps are fused to achieve a small number of parameters in the 2D network. Through the extraction of 3D features in the MRI data, D-UNet can achieve better performance than a pure 2D network.

\textbf{Enhanced Mixing Loss:} Second, in order to improve the convergence speed of the network, we propose a new loss function, called the Enhanced Mixing Loss, which not only enhances the gradient propagation of the traditional Dice Loss, but also combines the advantages of the Dice loss and Focal loss functions. This new method converges faster than using the two traditional loss functions, and exhibits a smoother convergence curve. In summary, this work has the following contributions: 
1. We propose the D-UNet network to effectively segment the lesion area in the MRI data. The structure is based on the 2D UNet improvement. A part of the 3D convolution is added to the downsampling module to extract the spatial information in the MRI volume data; the extracted features are fused with the 2D structures in a new method.
2. We propose a novel loss function, which is expected to make the network converge in a faster and smoother fashion. It would not only enhance the gradient propagation in the traditional Dice loss, but also combine the merits of Dice loss and Focal loss functions.
3. The proposed method is tested on the ATLAS dataset and compared to three state-of-the-art, demonstrating the superior performance of the method.

{\section{RELATED WORKS}

We summarize some of the work related to stroke segmentation, including hand-crafted feature based methods and deep learning based methods. Among them, the deep learning methods include 2D-based CNN, 3D-based CNN, and the traditional segmentation loss function.

Hand-crafted feature based methods: Researchers have been working on the automatic segmentation and prediction of brain disease areas and have achieved good results\cite{18}. Kemmling et al. \cite{19} use a multivariate computed tomography perfusion (CTP)-based model to calculate the probability of voxelwise infarcts. Kuo et al.\cite{20} propose to use the SVM classifier to learn texture feature vectors for the segmentation of liver tumors. Chyzhyk et al. \cite{21}propose to construct an image data classifier from multimodal MRI data for voxel-based lesion segmentations. Sivakumar et al. \cite{22} use an adaptive neuro fuzzy inference system (ANFIS) classifier to detect and segment brain stroke areas automatically while using the heuristic histogram equalization technique (HHET) to enhance the internal regions of the brain image. These proposals in the literature are machine learning models based on multiple linear regressions, relying on the precise design of features by feature engineers. They achieve good performance on small sized data sets, but have limited generalization in larger data sets.

Deep learning based methods: Deep learning has emerged in recent years, which address a key limitation in traditional machine learning methods, which require engineers to artificially design features. Chen et al. \cite{23} propose a 2D network framework consisting of an ensemble of a DeconvNets (EDD)-Net and a multi-scale convolutional label evaluation net (MUSCLE Net); this ensemble achieves the best performance on a large clinical dataset. Cui et al.  \cite{24} propose a network of cascaded structures for processing nasopharyngeal carcinoma cases in MRI images. The authors firstly segment the tumors and then classify the segmentation results to obtain four subregions of nasopharyngeal carcinoma. These deep learning methods convert the MRI data to 2D slices and apply 2D segmentation CNN for each slice. The 3D results are generated by connecting the 2D segmentation results. However, due to the limitations of the slices’ 2D characteristics, the important 3D context information in the volume data is neglected, thus the prediction may lose continuity.

3D CNN has proven to have great potential in the analysis of 3D MRI data. Kamnitsas et al. \cite{25} propose a two-path 3D CNN structure and uses a 3D fully connected conditional random field for post processing, which ranks first in the challenge of chronic stroke lesion segmentation (ISLES 2015). Zhang et al. \cite{10} propose 3D FC-DenseNet, which can make the network deeper by using the improved dense net tight connection structure to enhance the back propagation of image information and gradients. Feng et al. \cite{26} extract features from both the temporal and the spatial dimensions by using 3D convolution operations, which capture the dynamic information in multiple adjacent frames. However, they usually require more parameters and might sometimes over-fit on small training data sets \cite{27,28}. In addition, 2D-based and 3D-based cascade methods have emerged. For example, Li et al. \cite{29} propose a hybrid densely connected UNet (H-DenseUNet), which first performs a 2D-based dense-UNet segmentation, and then uses a 3D-based CNN to correct the spatial continuity of the liver and the tumor.

The binary cross-entropy loss function \cite{30} is commonly used in deep learning based segmentation tasks. This function calculates the gradient by characterizing the difference in the probability distribution of each pixel in the predicted sample and the real sample. Tsung-Yi Lin et al.\cite{31} add a modulating factor to deal with the serious imbalance between the number of foreground and background pixels. Another common loss function is Dice's coefficient loss \cite{32}. This function directly calculates the gradient by the dice overlap coefficient of the predicted sample and the real label; it can also alleviate to some extent the segmentation problem resulting from the pixel imbalance between the foreground and the background.

{\section{Methods}
In this section, we introduce our approach including the proposed D-UNet framework, enhanced mixing loss, and the implementation details. In Section 3.1, we illustrate the proposed D-UNet framework, in Section 3.2, we introduce the enhanced mixing loss algorithm, and finally in Section 3.3, we present some implementation details.

\begin{figure*}[htbp]
	\centering
	\includegraphics[width=\linewidth,scale=1]{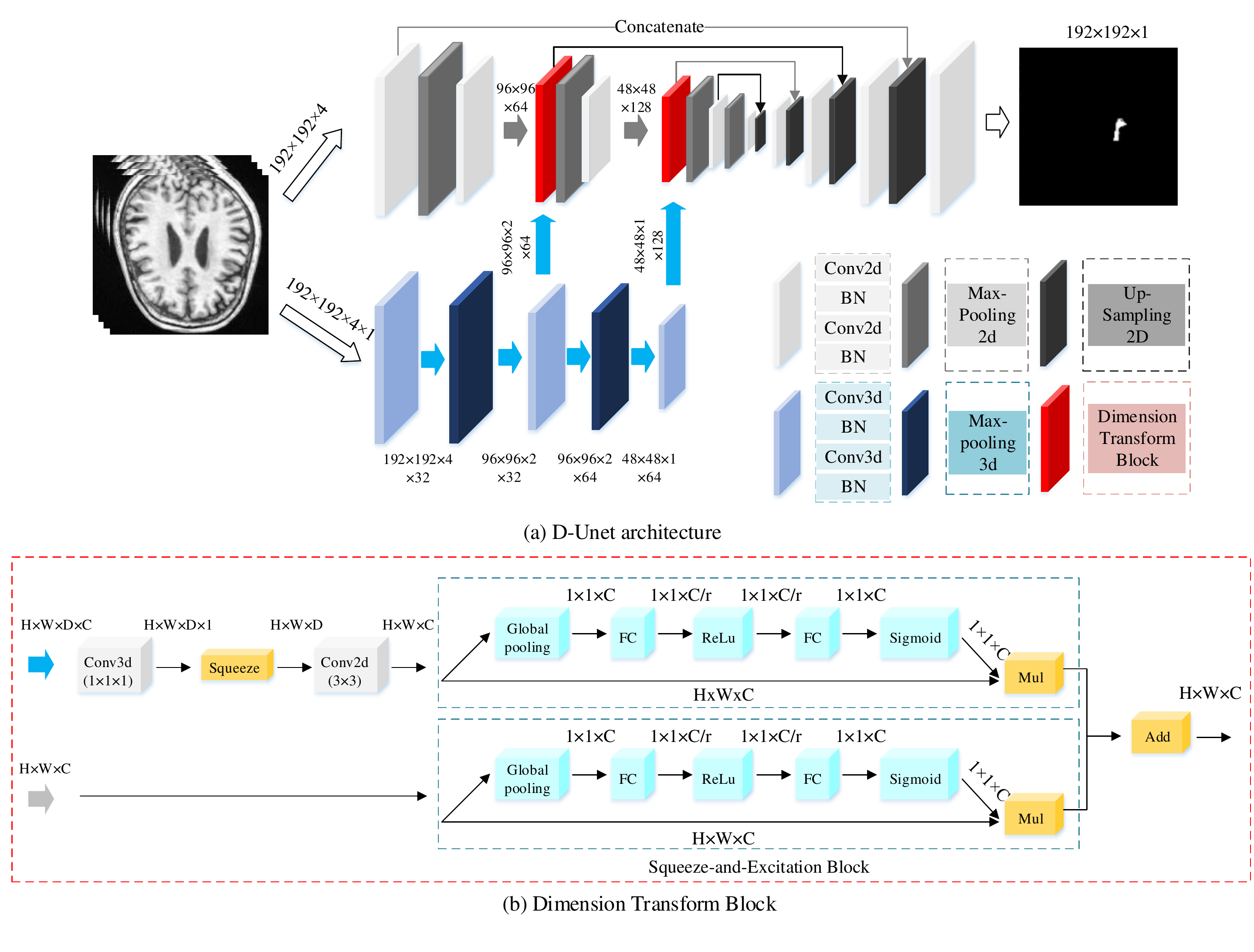}
	\caption{The entire D-UNet architecture is shown in (a). This network improves 2D UNet, which combines 3D convolution in the downsampling phase and uses a dimension transform block to combine them. (b) Introduces the details of the dimension transform block, which has two branches for its input from 2D and 3D networks. First, the feature channel of the 3D network output (blue arrow) is compressed to 1 by using a 1x1x1 convolution. The compressed result is then squeezed in a spatial dimension and passed to a 2D 3x3 convolution. This makes the output consistent with the 2D network (gray arrow). Finally, each of the channels is weighted by the SE-block and then added together.} \label{fig2}
\end{figure*}

\subsection{D-UNet for extracting three-dimensional information}
The basic structure of the network consists of an improved UNet \cite{33}. This symmetrical encoder-decoder structure combines high-level semantics with low-level fine-grained surface information; it has achieved good effects on medical images. The encoding phase of the D-UNet consists of two dimensions. As shown in Fig. 2(a), both the 2D and 3D convolutions perform the downsampling operation in their respective dimensions; the results are combined through the dimension transform block which denote as a red cube. This fusion enables subsequent 2D networks to be integrated into the 3D information, refines the edges of the target area, and facilitates the ability of the network to identify small lesion areas. Meanwhile, since the 3D information is well extracted in the early stage of the network, and the trainable parameters of the network are extremely increased as the network deepens, the dimension transform block is only added in the early coding stage.

Specifically, consider Fig. 2(b), where H$\times$W denotes the feature dimensions of height and width, D represents the depth in the volume feature, and C represents the channel of the feature map. The dimension transform block consists of 3D dimensionality reduction, channel excitation \cite{34}, and dimensional fusion. The squeeze-and-excite (SE) block has been proposed in recent years, where $r$ denotes the reduction ratio, a hyperparameter which allows us to vary the capacity and computational cost of the SE block [34]. This block activates the connection between different channels by weighting the feature channels. We apply this structure in the dimension fusion block in order to enhance the fusion effect of 3D features and 2D features. 

In each dimension transform block, we first reduce the dimensions of the 3D branch feature map and then add with the 2D branch after SE weighted respectively. Specifically, let $I_3d$ and $I_2d$ denote the feature maps from 3D and 2D network respectively, which act as the input of dimension transform block, n denotes the batch size, $h\times w\times d$ denotes the maps’ height, width, depth and last dimension c denotes the maps’ channel. We first convert I$_{3d}$\(\in\)R$^{n\times h\times w\times d\times c}$ to I$^{*}$$_{3d}$\(\in\)R$^{n\times h\times w\times d\times 1}$ by using a 3D $1\times1\times1$ convolution which filter number is set to 1, then we squeeze the dimensionality of  I$^{*}$$_{3d}$ from $n\times h\times w\times d\times 1$ to $n\times h\times w\times d$. In order to keep the channel number consistent with the 2D branch for later integration, we also convert I$^{*}$$_{3d}$\(\in\)R$^{n\times h\times w\times d}$ to I$^{*}$$_{3d}$\(\in\)R$^{n\times h\times w\times c}$ by using a 2D 3$\times$3 convolution that filter number is set to c. Let I$^{'}$$_{3d}$ denote the I$_{3d}$ after dimensionality reduction:
\begin{equation}
I^{'}_{3d}=f_{r}(I_{3d}),  I^{'}_{3d}\in R^{n\times h\times w\times c}
\end{equation}
where f$_{r}$ indicates the dimensionality reduction operation, thus we convert the size of 3D feature map from I$_{3d}$\(\in\)R$^{n\times h\times w\times d\times c}$ to $I^{'}_{3d}\in R^{n\times h\times w\times  c}$. In order to enhance the feature expression ability of the two dimensions before fusion, we use an SE block to weight the 3D and 2D feature map channels, and add their channel weighted outputs:

\begin{equation}
T=f_{SE}(I^{'}_{3d})+f_{SE}(I_{2d}),        T \in R^{n\times h\times w\times c}
\end{equation}

The 3D and 2D features are fused in this step, where f$_{SE}$ denotes the SE weighted block proposed in \cite{34}. $T$ denotes the feature map fusion which results in the dimension fusion block. More detailed parameter settings for the entire network are shown in Table 1.
\begin{table*}[tp]  
	\fontsize{6}{6}\selectfont  
	\centering  
	\caption{Architecture of the proposed D-UNet.
		‘Feature size’ denotes the size of the manipulated feature map while the last dimension indicates the channel number. Up-sampling – [*] indicates that the corresponding layer number is concatenated before up sampling and $N^{*}$ indicates the number of features of the corresponding layer number, for example, Up-sampling block 1 being connected to Convolution block 4, $N$ is set to 256, up sampling block 2 being connected to dimension fusion block 3, $N$ is set to 128, and so on.}  
	\label{tab1}
	\begin{tabular}{ccccc}  
		\toprule  		
		& \bf Feature size & \bf Two-dimensional operation&\bf Feature size&\bf Three-dimensional operation\cr 
		\midrule  
		\bf Input					& 192$\times$192$\times$4& -& 192$\times$192$\times$4$\times$1& -\cr 
		\midrule  
		\bf Convolution block 1		& 192$\times$192$\times$32& 2$\times$(3$\times$3 Conv+ Bn)& 192$\times$192$\times$4$\times$3&  2$\times$(3$\times$3$\times$3 Conv+ Bn)\cr 
		\midrule  
		\bf Pooling					& 96$\times$96$\times$32& 2$\times$2  max pooling& 96$\times$96$\times$2$\times$32& 2$\times$2$\times$2  max pooling\cr 
		\midrule  
		\bf Convolution block 2		& 96$\times$96$\times$64& 2$\times$(3$\times$3 Conv+ Bn)& 96$\times$96$\times$2$\times$64& 2$\times$(3$\times$3$\times$3 Conv+ Bn)\cr 
		\midrule  
		\bf Dimension fusion block 2& 96$\times$96$\times$64& -& -& -\cr 
		\midrule  
		\bf Pooling					& 48$\times$48$\times$64& 2$\times$2  max pooling& 48$\times$48$\times$1$\times$64& 2$\times$2$\times$2 max pooling\cr 
		\midrule  
		\bf Convolution block 3		& 48$\times$48$\times$128& 2$\times$(3$\times$3 Conv+ Bn)& 48$\times$48$\times$1$\times$128& 2$\times$(3$\times$3$\times$3 Conv+ Bn)\cr 
		\midrule  
		\bf Dimension fusion block 3& 48$\times$48$\times$128& -& -& -\cr 
		\midrule  
		\bf Pooling					& 24$\times$24$\times$128& 2$\times$2  max pooling& -& -\cr 
		\midrule  
		\bf Convolution block 4		& 24$\times$24$\times$256& 2$\times$(3$\times$3 Conv+ Bn)& -& -\cr 
		\midrule  
		\bf Dropout					& 24$\times$24$\times$256& -& -& -\cr 
		\midrule  
		\bf Pooling					& 12$\times$12$\times$256& 2$\times$2  max pooling& -& -\cr 
		\midrule  
		\bf Convolution block 5		& 12$\times$12$\times$512& 2$\times$(3$\times$3 Conv+ Bn)& -& -\cr 
		\midrule  
		\bf Dropout					& 12$\times$12$\times$512& -& -& -\cr 
		\midrule  
		\bf Up-sampling block 1-4	& 192$\times$192$\times$32& 2$\times$2 Up-sampling–[*] &-&- \\  & & \ \ 2$\times$(3$\times$3 Conv+ Bn)\cr 
		\midrule  
		\bf Convolution				& 192$\times$192$\times$1& 1$\times$1 Conv& -& -\cr 
		
		\bottomrule  
	\end{tabular}  
\end{table*}

\subsection{Enhanced Mixing Loss Function}
In 3D medical data, especially MRI stroke images as shown in Fig. 1, the volume occupied by the stroke is often very small throughout the scan interval. An extremely large number of background regions may dominate the loss function during training, which leads to the learning process easily falling into a local optimal solution. Therefore, we propose a new loss function, which refers to the method addressing the foreground-background voxel imbalance in \cite{31}, and combines two traditional loss functions in a concise manner\cite{32}.

\subsubsection{Focal Loss}
Focal loss (FL) is an improvement of the binary cross entropy loss (BCE), by adding a modulating factor. This reduces the loss contribution from easy samples and extends the range in low loss. We introduce the formula of focal loss from the binary cross entropy (BCE):

\begin{eqnarray}
FL(p,g) =
\begin{cases}
-\sum_{i=1}^{N_{f}}\alpha(1-p)^{\gamma}\log(p), \ \ \ \ \ \ \ \ \ if \ g = 1 \\
-\sum_{i=1}^{N_{b}}(1-\alpha)p^{\gamma}\log(1 - p), \ \ \   otherwise
\end{cases}
\end{eqnarray}

where $g\in{0,1}$ represents the ground truth based on the pixel level; $p\in[0,1]$ represents the model prediction probability value, in which 0 denotes the background and 1 is the foreground; $N_f$ and $N_b$ represent the numbers of pixels of class 0 and class 1, respectively;  $\alpha\in(0,1]$ and $\gamma\in[0,5]$ are the modulation factors, which can be flexibly adjusted according to the situation. 

\subsubsection{Dice Coefficient Loss}
The dice coefficient loss (DL) mitigates the imbalance problem of background and foreground pixels by modifying the segmentation evaluation index DSC between the prediction samples and the ground truth annotation, showing better performance in the segmentation task:

\begin{equation}
DL(p,g) = 1- \frac{2\sum_{i=1}^{N}p_{i}g_{i}+\delta}{\sum_{i=1}^{N}p^{2}_{i}+\sum_{i=1}^{N}g^{2}_{i}+\delta}
\end{equation}
where $\delta\in[0,1]$ is a tunable parameter to prevent a divide-by-zero error and let the negative samples also have a gradient propagation.

\subsubsection{Proposed Enhanced Mixing Loss}
Based on the above two kinds of loss, we propose the enhanced mixing loss (EML) to increase the convergence speed. First, Log value was used in DL and we invert the value for keeping the value positive, thus enhancing the gradient obtained for each iteration. Then, in order to explore whether the two losses have mutually reinforcing relationships, we also add the focal loss. However, since the focal loss is based on the sum of all voxel probabilities, it is numerically much larger than the dice loss ($DL(p,g)\in[0, 1]$), which plays a leading role in gradient propagation. We hope that the newly added FL and log(DL) contribute equally to EML, so a balance factor of $1/N$ is added to FL to obtain a FL based voxel average. The formula for EML is as follows:
\begin{equation}
EML(p,g)=  \frac{1}{N}FL(p,g)-\log(DL(p,g))
\end{equation}

\subsection{Implementation details}
In the data preprocessing, transverse section images have been selected for this experiment. Within each image, a square area is selected with the diagonal coordinates (10, 40) and (190, 220). This selected area eliminates irrelevant information and enlarges the proportion of stroke lesions in the entire image. Next, the cropped images are resized  $192\times192$ using a bilinear interpolation. Finally, each slice of the processed image is integrated, with a spatial arrangement of two upper slices and one lower slice, forming a matrix of size $192\times192\times4$.
In the downsampling phase, modifications to reduce the total number of parameters from UNet are made. Specifically, the number of filters in the first convolution in the 2D-based and 3D-based streams is set to 32. After each pooling layer, the number of convolution filters is doubled, and finally the number of convolutions in the 2D stream is set to 512. The kernal initialization for each convolution is set using the He’s method \cite{35}. A batch normalization is conducted after each layer of convolution to improve the stability of the training. The parameter r in the dimension transform block is set to 16. $\alpha,\gamma,\delta$ in the loss function is set to 1.1, 0.48, 1 respectly to fit our randomly selected dataset. With the SGD optimizer, the learning rate is set to 1e-6. Additional parameter settings are consistent with\cite{36}. We have also employed data augmentation methods to improve the robustness of the model, including setting the input mean to zero translation, scaling, and horizontal flipping. These methods are applied to all of our comparative experiments to ensure fairness. We have trained the models on three 1080TI GPUs. All of the models are trained using the first 150 epochs before validating, to optimize the performance of each architecture without any fine tuning.

\section{EXPERIMENTAL RESULTS AND DISCUSSIONS}
We compare our proposed method to the 2D and 3D convolutional UNet. In this section, we also show the superiority of the proposed loss and discuss the results of the proposed dimension-transform block at different stages.
Datasets and quantitative indicators: We have used the Anatomical Tracings of Lesions-After-Stroke (ATLAS) dataset \cite{7} as our training and validation sets. The dataset contains 229 cases of chronic stroke with MRI T1 sequence scans, in which the size of each case is 233$\times$197$\times$189 while the physical size is 0.9$\times$0.9$\times$3.0mm$^3$; the scans delineate different lesion grade staging. We have randomly selected 183 cases (accounting for the overall 0.8 ratio) as the training set, and the remaining cases as validation sets. We report the model's performance in the Dice Similarity Coefficient (DSC), precision, and recall. DSC is an important indicator to assess the overall difference between our estimates and the ground truths. Recall usually reflects the extent of recall in the lesion area, which is an important reference in clinical practice. We perform threshold processing on all of the prediction results. When the probability that the pixel is predicted to be foreground is less than 0.5, we set it to zero, otherwise it is set to one. In addition, precision evaluates the quality of the segmentation, as the proportion of boundary pixels in the automatic segmentation that correspond to boundary pixels in the ground truth of the image. The quantitative indicator formulae are shown below:
\begin{equation}
DSC=  \frac{2TP}{2TP+FP+FN}
\end{equation}

\begin{equation}
Recall=  \frac{TP}{TP+FN}
\end{equation}

\begin{equation}
Precision=  \frac{TP}{TP+FP}
\end{equation}

where, true positive $(TP)$ indicates that the model correctly predicted voxel. False positive $(FP)$ indicates the voxel that the model classify negative as positive. False negative $(FN)$ indicates that the positive voxel is mistakenly classified as negative by the model.

\subsection{Performance comparison with UNet and its 2D and 3D variants}
We compare the proposed method with the baseline architecture, UNet, both qualitatively and quantitatively. We have trained two versions of UNet to better exploit its performance. The original version uses the parameter settings described in \cite{33}. In the transformed version, the number of convolution kernels is reduced and batch normalization is added after each convolution. Section 3.3 provides a more detailed description of the transform. Furthermore, we compare the proposed method with the 3D structure to prove that our method can extract the 3D information of the data with a small amount of structure. We show the results of different structure outputs, and choose case based DSC, recall, and precision as our quantitative indicators. Furthermore, we also conduct a global based DSC assessment as an auxiliary judgment. It is worth noting that the batch size of all networks is set to 36 during training, except for the 3D UNet. Since the 3D Unet has very high memory demands, the batch size is set to six.

In the quantitative comparison, as shown in Table 2, the UNet (original), UNet (transform), and the proposed method are evaluated. The 2D ‘transform’ is close to the ‘original’ in both indicators. However, with respect to the kernel number, the transformation scheme is only half of the original network. It shows that a large number of convolution kernels are unnecessary for this small data set. We have also reduced the number of 3D Unet convolution kernels to match the number of 2D Unet convolution kernels. In addition, all the transform versions and our proposed method adds batch normalization after each convolution layer (except for the last layer of convolution) to ensure convergence stability. The 2D structure achieves a better index than the 3D structure, we consider the following factors: first, due to its huge network structure, 3D network requires a huge amount of time to train, making it difficult to adjust the hyperparameters to optimize performance. Then, Overfitting can easily occur in large network structures, especially if there are fewer training sets (our training set contains only 183 cases). Last but not least, 3D structure requires a large amount of computing resources, which limits the width and depth of the 3D structure by our implementation platform conditions, leading to a decrease in the performance. The results we reported is also consistent with that of \cite{29}, which also shows the phenomenon that the segmentation performance of the 3d structure is lower than the 2d structure. In addition, the time required to train a 3D UNet (140 hours) is about six times that of 2D UNet (23.8 hours). By comparing the segmentation performance and the number of parameters, the metrics for the proposed method are improved by 3.83\% over a single 2D structure, and the number of parameters is only increased by 2\%. These results show that with the addition of 3D structure, the overall performance of the network is improved. It proves that compared with the simple 2D network structure, the proposed dimension transform block can effectively utilize 3D information.

\begin{table}[tp]  
	\fontsize{4}{10}\selectfont  
	\centering  
	\caption{ Comparison results of the proposed method with baseline approaches.}  
	\label{tab2}
	\begin{tabular}{c|ccccc}  
		\toprule  		
		\bf Method&\bf \ \ DSC\ \ &\bf DSC(global)&\bf \ \ Recall\ \ &\bf \ \ Precision\ \ \bf&\bf \ \ Total parameters\ \ \cr 
		\midrule  
		2D UNet(original)&0.4874$\pm$0.2858&0.7117&0.4838$\pm$0.2983&0.5612$\pm$0.3229&31,030,593\cr  
		2D UNet(transform)&0.4966$\pm$0.2906&0.7146&0.5038$\pm$0.3044&0.5511$\pm$0.3298&\bf7,771,297\cr  
		3D UNet(transform)&0.4710$\pm$0.2877&0.7098&0.4736$\pm$0.3099&0.5531$\pm$0.3247&22,597,826\cr  
		Ours			   &\bf0.5349$\pm$0.2763&\bf0.7231&\bf0.5243$\pm$0.2910&\bf0.6331$\pm$0.2958&8,640,163\cr  
		\bottomrule  
	\end{tabular}  
\end{table}

\subsection{Comparison with other state-of-the-art methods}
In this section, we compare the proposed model with other existing well-known frameworks and methods mentioned in \cite{37}. SegNet \cite{38} performs nonlinear upsampling; this is accomplished  using unpooling for the maxpooling index defined in the downsampling phase. Pyramid Scene Parsing Network \cite{39} introduces more context information by using atrous convolution and pyramid pooling. DeepLab v3 plus \cite{40} is the latest version of the Google DeepLab series, which combines the advantages of Atrous Spatial Pyramid Pooling and an encoder-decoder structure. The above methods have achieved good results in the deep learning segmentation task. We compare with these methods to demonstrate the superior performance of the proposed D-UNet. All of the network parameters are configured according to the original articles, except for the minor changes mentioned in Section 4.1; the networks are implemented in our platform. The training set and validation set are strictly consistent (including data preprocessing). During training, the batch size is set to 36. Additional details can be viewed the open source code.

The prediction results obtained with the different frameworks are illustrated in Fig. 3. Seven prediction maps are randomly selected and sorted by the size of the lesion area in ascending order. As shown in the figure, it can be observed that the lesion with a very small foreground area is quite fuzzy, indicating this case is very difficult for the network. In spite of this, our method correctly detects the lesion area. It proves that our approach possesses the ability to identify difficult samples. From the third and fourth rows of Fig. 3, we have found that all models correctly predicted the location of the lesion area in the case of big differences in foreground and background. Furthermore, our method is closer to the ground truth in terms of lesion boundaries. Observing the last two rows, we have found that the proposed method still maintains a good, feature expression ability when the boundary is very blurry. This is because we combine the 3D features and the spatial dimensions to more effectively express the characteristics of the edge blurred lesions.
\begin{figure*}[htbp]
	\centering
	\includegraphics[width=\linewidth,scale=1]{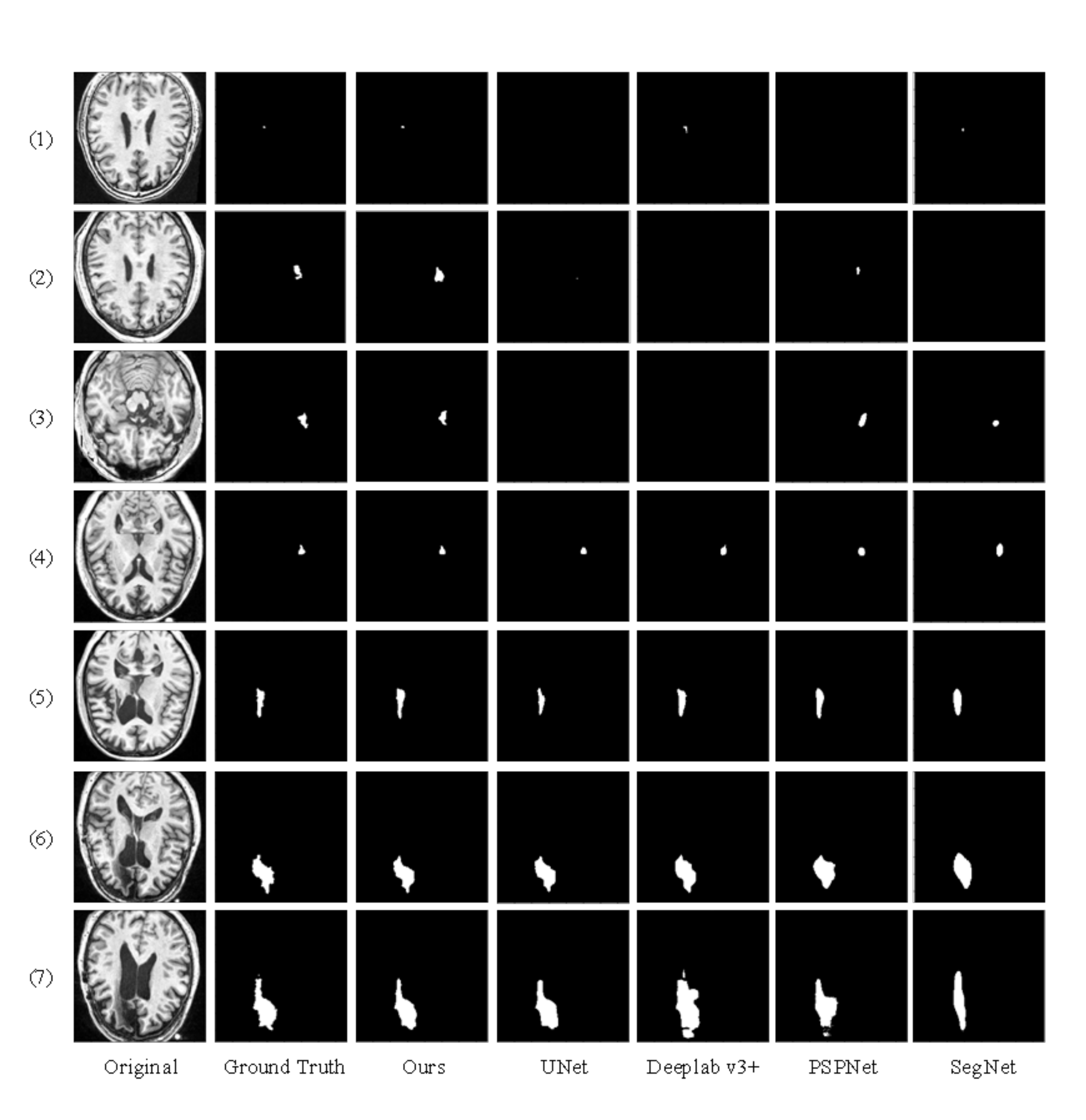}
	\caption{Comparisons of our method, Baseline, DenseUnet, DeepLabv3+, PSPNet, and FCN-8s on four different patients.} \label{fig3}
\end{figure*}
In order to show the distribution of the results and prove the stability of the proposed method, we draw a box plot based on the DSC score for each case. These results are shown in Fig. 4. The lower edge of all methods is zero because the ATLAS data set has a very large number of small lesion areas (e.g., the first to third rows of Fig. 3), which easily cause the model to fail to recognize the lesions. The upper edge and median line for the proposed method are the best in comparison with other highly recognized methods, which means it not only achieves the highest segmentation performance on a single case, but also yields better median scores for all cases.

\begin{figure}[htbp]
	\centering
	\includegraphics[width=\linewidth,scale=1]{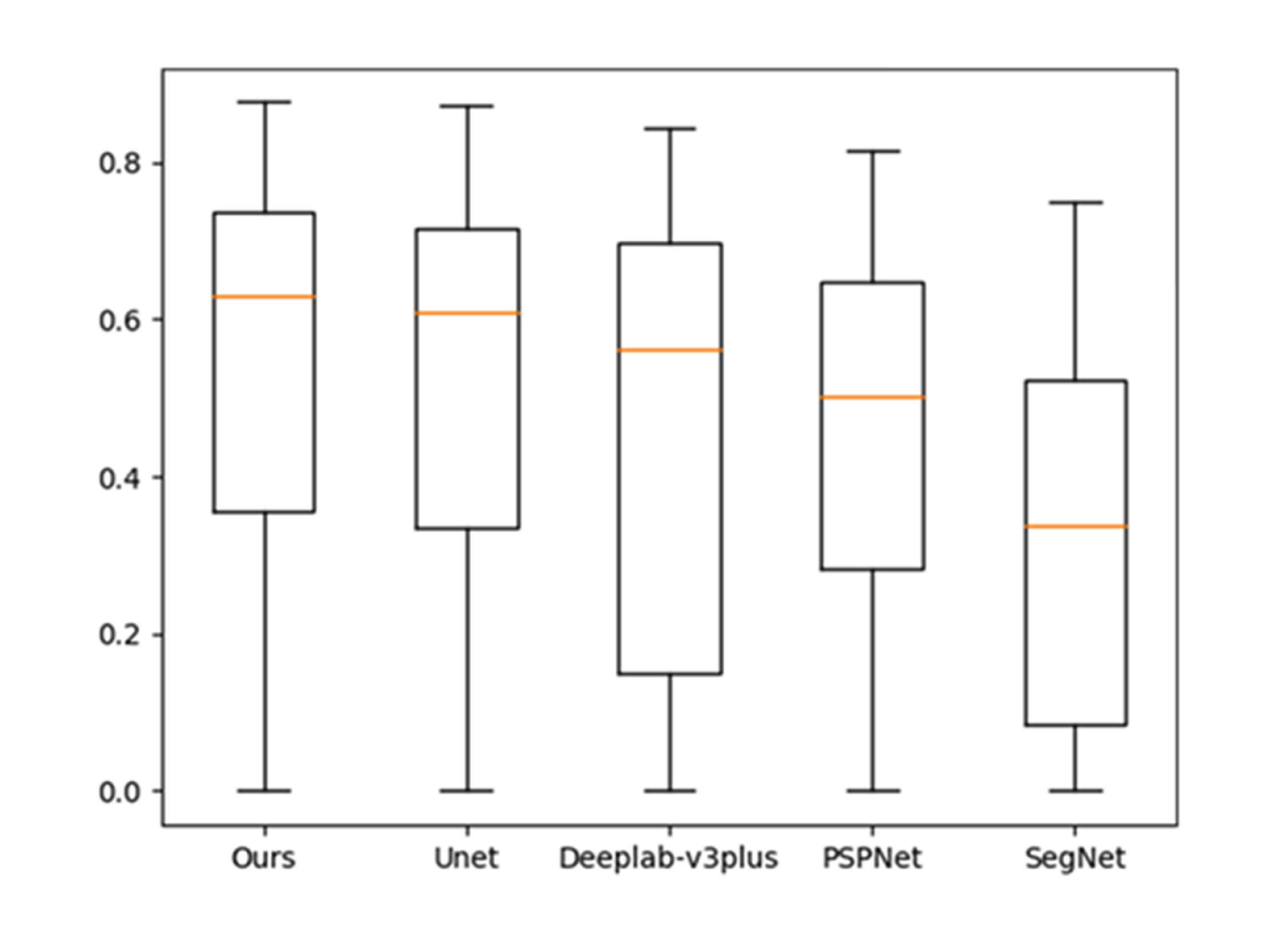}
	\caption{Box plots of DSC score results for different methods.} \label{fig4}
\end{figure}

To quantitatively illustrate the superiority of our method, we compare the results of these algorithms and summarize them in Table 3, where DSC, recall, and precision are based on the mean $\pm$ standard deviation of each case, and DSC (global) represents the metric based on the voxel calculation. Hu et al. use the ATLAS dataset, and select some specific cases as training/validation sets to summarize some traditional segmentation methods. We compare several deep learning segmentation frameworks with the methods mentioned in \cite{37}. The top half of the table shows the implementation in\cite{37}, and the bottom half is implemented on our platform. It can be seen from the table that the deep learning based methods achieve better performance than the traditional algorithms. With respect to the DSC scores, the proposed method ranked first with a score of 0.7231. This is 0.30 higher than the Clusterize method in DSC (the lowest), and 0.0383 higher than the UNet method (second best). Our method is superior in terms of the segmentation performance for each case. DSC (global) can reflect a voxel-based overall DSC score more intuitively. With respect to recall and precision, the performance of the traditional algorithm Clusterize is the highest in recall, but its precision score is the lowest, which indicates that this algorithm identifies many non-lesional areas as lesions, thus causing a high recall. The proposed method ranked third in the recall score of 0.5243, and the highest in the precision score of 0.6631, which indicates that the identified regions are basically the correct lesion areas.
\begin{table}[tp]  
	\fontsize{5}{10}\selectfont  
	\centering  
	\caption{The quantitative comparison of different methods.
		Among them, DSC, Recall, and Precision are based on the mean $\pm$ standard deviation calculated in each case, and DSC (global) represents the DSC based on the voxel calculation.
	}  
	\label{tab3}
	\begin{tabular}{c|cccc}  
		\toprule  		
		\bf Method&\bf \ \ DSC\ \ &\bf DSC(global)&\bf \ \ Recall\ \ &\bf \ \ Precision\ \ \bf\cr 
		\midrule  
		Clusterize				&0.23$\pm$0.19		&-		&\bf0.79$\pm$0.23	&0.16$\pm$0.15\cr  
		ALI						&0.36$\pm$0.25		&-		&0.55$\pm$0.31		&0.31$\pm$0.25\cr  
		Lesion gnb				&0.36$\pm$0.23		&-		&0.69$\pm$0.29		&0.30$\pm$0.20\cr  
		LINDA			   		&0.45$\pm$0.31		&-		&0.52$\pm$0.34		&0.50$\pm$0.34\cr  
		\midrule  
		SegNet					&0.3292$\pm$0.2514		&0.5993		&03318$\pm$0.2654		&0.3846$\pm$0.2883\cr  
		PSP						&0.4462$\pm$0.2633		&0.6729		&0.4704$\pm$0.2780		&0.4998$\pm$0.2913\cr  
		Deeplab v3 plus			&0.4529$\pm$0.2921		&0.7104		&0.4456$\pm$0.3032		&0.5627$\pm$0.3249\cr  
		UNet			   		&0.4966$\pm$0.2906		&0.7146		&0.5038$\pm$0.3044		&0.5511$\pm$0.3298\cr  
		Ours			   		&\bf0.5349$\pm$0.2763	&\bf0.7231	&0.5243$\pm$0.2910		&\bf0.6331$\pm$0.2958\cr  
		\bottomrule  
	\end{tabular}  
\end{table}

\subsection{Loss validity}
In order to compare the effectiveness of the proposed loss, we trained on the proposed model and compared several common losses in the segmentation task. As illustrated in Fig. 5, the results show several DSC rising graphs on the training sets as the number of training iterations increases. Comparing the scores of the losses in the training set, in the early stage of training (about 30 epoch) shown on Fig. 5, the dice coefficient loss converges faster than the focal loss, but is subsequently exceeded by the focal loss. In each period, our method converges faster than other methods.
\begin{figure}[htbp]
	\centering
	\includegraphics[width=\linewidth,scale=1]{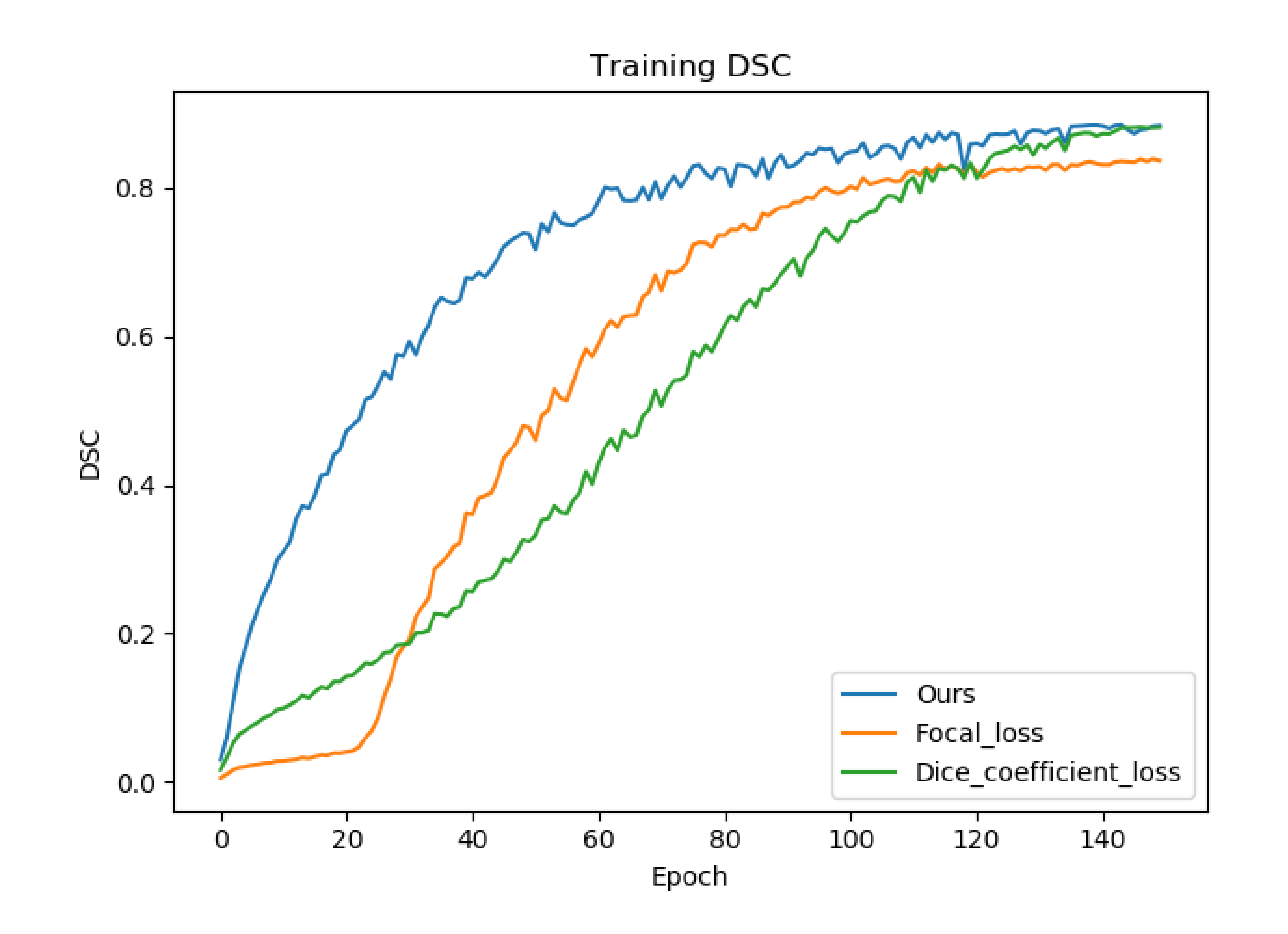}
	\caption{The DSC score curve of the training set during the training process.} \label{fig5}
\end{figure}

We also performed a quantitative analysis of the three losses, as shown in Table 4. It is worth noting that our goal is to make the proposed model converge faster. This statistical score is only used as a secondary reference. Since DL directly uses 1 - DSC as punished, it has an advantage to achieve the highest DSC score. Therefore, despite the proposed EML is slightly lower than DL in DSC (0.0115 lower) and Recall (0.057$\pm$0.002 lower), we consider that it is within an acceptable range. On the other hand, EML achieves the highest precision whereas DL presents the lowest, which indicates that EML can effectively reduce the false positive results (This phenomenon is also consistent with FL). Therefore, EML is competitive in segmentation performance compared to DL and FL.
\begin{table}[tp]  
	\fontsize{8}{10}\selectfont  
	\centering  
	\caption{Quantitative analysis of the three loss functions.}  
	\label{tab4}
	\begin{tabular}{c|ccc}  
		\toprule  		
		\bf Method&\bf \ \ DSC(global)\ \ &\bf \ \ Recall\ \ &\bf \ \ Precision\ \ \bf\cr
		\midrule  
		FL &0.6805&0.4339$\pm$0.2625&0.6225$\pm$0.3667\cr  
		DL &\bf0.7346&\bf0.53$\pm$0.2908&0.6143$\pm$0.3324\cr  
		EML&0.7231&0.5243$\pm$0.2910&\bf0.6331$\pm$0.2958\cr  
		\bottomrule  
	\end{tabular}  
\end{table}

\subsection{Comparison of dimension fusion blocks between different layers}
We also compare the results of the dimension transform block used in different ways (Add, SE) and in different downsampling layers. The DSC is used as the main evaluation index of model performance, and the number of their parameters is enumerated. The 3D framework used in the experiment, as shown in Fig. 2, has only two pooling layers for the dimension transform block within three layers. The 'add' in Table 5 represents the use of the final fusion operation of the dimension fusion block, (i.e., the Add block in Fig. 2), and the 'SE' indicates the use of the SE block the figure \cite{34}. The last column of the name in the architecture indicates which layer is used in the conversion structure. For example, 'Add-23' means the corresponding 3D structure fusion before the second and third maxpooling in the 2D structure.

In order to prove the validity of dimensional transformation on 2D networks, we compare the performance of UNet (division) and different structures using the dimension fusion block in different layers. All of the results with using the dimensional fusion are better than UNet without using the dimension fusion block. We suspect this is because the network fuses the 3D features in downsampling. By comparing ‘Add-1’, ‘Add-12’, and ‘Add-23’, we find that the deeper the 3D structure is, the better the fusion performance results. This can be interpreted as a deeper 3D structure provides a better feature extraction ability. However, we found an interesting phenomenon: the DSC with more layers in the case ‘Add-123’ has decreased. This may be because the gradient of the 3D and 2D structure is tighter with the fusion of more layers; this change in gradient may result in a decrease in the efficiency of the feature extraction. We demonstrate the validity of the proposed fusion structure and demonstrate that our method obtains 3D information in an efficient manner.

To prove that the SE block can achieve better fusion by weighting the output of the two dimensions, we show a comparison of the addition of SE (the last three rows) and direct fusion in Table 5. The structure of the SE block is consistent with the direct fusion for the number of layers, but all of the structures using the SE block are higher than the DSC of the direct fusion structure. It is shown that the nonlinear weighting before feature fusion can enhance the fusion effect of features.

\begin{table}[tp]  
	\fontsize{8}{10}\selectfont  
	\centering  
	\caption{ Comparison results of the proposed method with baseline approaches.}  
	\label{tab5}
	\begin{tabular}{c|cc}  
		\toprule  		
		\bf Architecture&\bf \ \ DSC\ \ &\bf \ \ Total parameters\ \ \cr 
		\midrule  
		2D UNet(transform)	&0.4966$\pm$0.2906&\bf7,771,297\cr  
		Add-1				&0.5102$\pm$0.2932&7,802,210\cr  
		Add-12				&0.5216$\pm$0.2776&7,970,019\cr  
		Add-23 				&0.5248$\pm$0.2770&8,635,043\cr
		Add-123				&0.5110$\pm$0.2762&8,636,260\cr  
		SE–Add-12			&0.5235$\pm$0.2851&7,971,299\cr  
		SE–Add-23			&\bf0.5349$\pm$0.2763&8,640,163\cr  
		SE–Add-123			&0.5186$\pm$0.2865&8,647,012\cr  
		\bottomrule  
	\end{tabular}  
\end{table}

\section{Conclusion}
Automated stroke segmentation plays an important role in clinical diagnosis and prognosis. It is of great value to quickly and accurately identify areas of the lesions and help physicians make surgical plans without high computing resource demands. In this paper, we propose an end-to-end training method for the automatic stroke segmentation, in which 3D context information can be effectively utilized, with low hardware requirements. Meanwhile, we propose a new loss function for faster and smoother convergence. The proposed method has been compared with three state-of-the-art methods; it achieves the best performance on two quality metrics (DSC = 0.5349$\pm$0.2763, Precision = 0.6331$\pm$0.2958). 
In future work, we hope to increase the punishment for extremely difficult samples inspired from \cite{41}, which may further enhance the performance of the proposed EML. We plan to validate our method on a larger clinical dataset to verify the generalization of the method in the current 3D structure, further study the possibility of dimension fusion block combinations, and extend our model to different applications, for example, calculating overlap of the lesion with existing brain regions or structures, for used as predictors of long-term stroke recovery and rehabilitation.

\ifCLASSOPTIONcompsoc
  \section*{Acknowledgments}
	This research was partly supported by the National Natural Science Foundation of China (61601450, 61871371, 81830056), Science and Technology Planning Project of Guangdong Province (2017B020227012, 2018B010109009), the Basic Research Program of Shenzhen (JCYJ20180507182400762), Youth Innovation Promotion Association Program of Chinese Academy of Sciences (2019351).
\else
  \section*{Acknowledgment}
\fi

\ifCLASSOPTIONcaptionsoff
  \newpage
\fi



%

%

\begin{IEEEbiography}[{\includegraphics[width=1in,height=1.25in,clip,keepaspectratio]{zhouyongjin-eps-converted-to}}]{Yongjin Zhou}
Yongjin Zhou (M'13) received the B.Sc., M.Eng and Ph.D. degrees in biomedical engineering from Xi'an Jiaotong University,Xi'an, China, in 1996, 1999, and 2003 respectively. He is currently an Associate Professor and the Head of the Department of Medical Electronics with the Shenzhen University, Shenzhen, China. His research interests include biological signal processing, medical image analysis, and pattern recognition.
\end{IEEEbiography}

\begin{IEEEbiography}[{\includegraphics[width=1in,height=1.25in,clip,keepaspectratio]{huangweijian-eps-converted-to}}]{Weijian Huang}
Weijian Huang received the B.E. degree in Shenzhen University, Shenzhen, China, in 2017. He is now studying for a master's degree in Shenzhen University, institutes of biomedical engineering. His current research interests include deep learning and computer vision.
\end{IEEEbiography}

\begin{IEEEbiography}[{\includegraphics[width=1in,height=1.25in,clip,keepaspectratio]{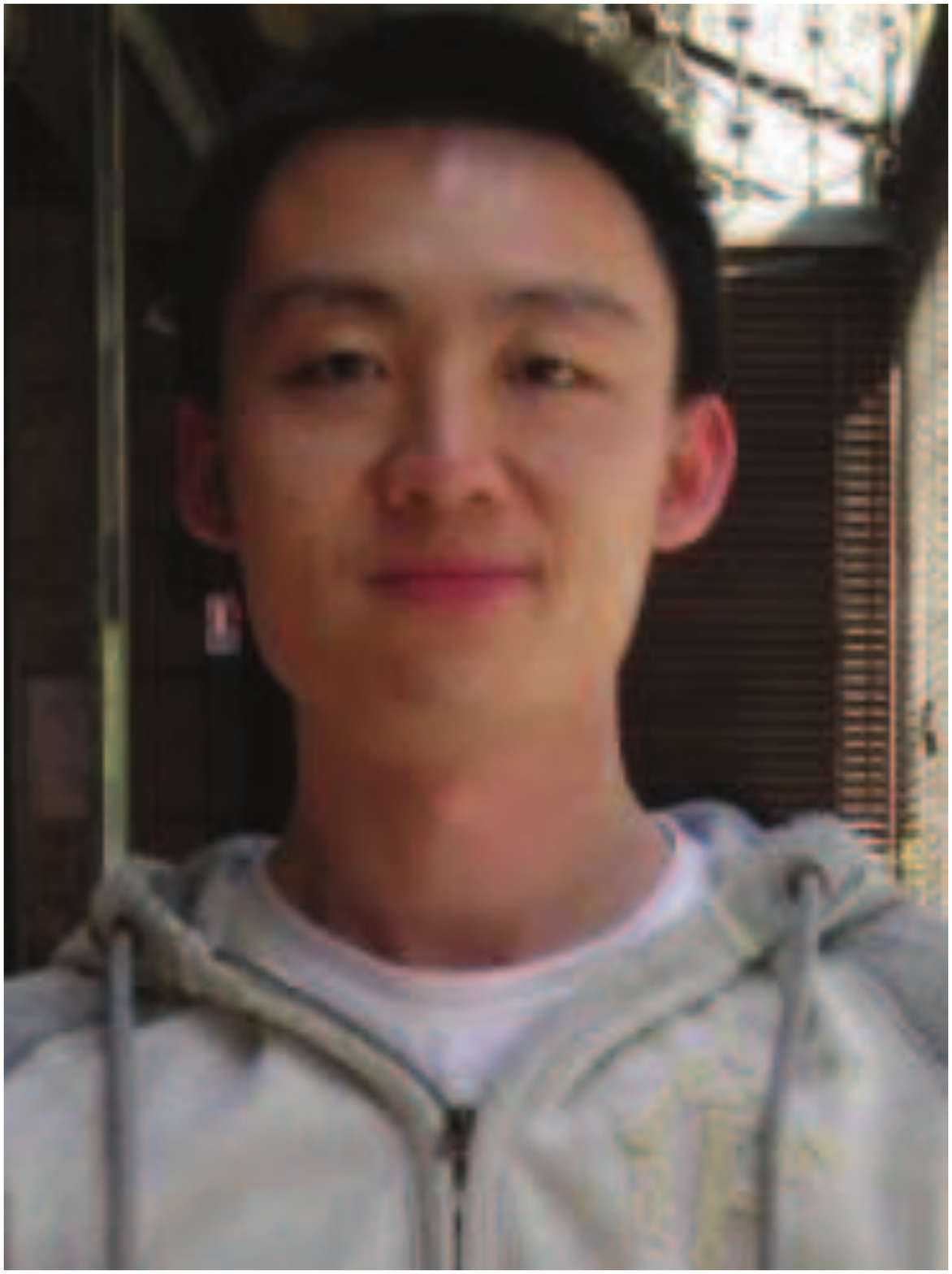}}]{Pei Dong}
	Pei Dong (S’12–M’15) received the B.E. degree in electronic engineering and the M.E. degree in signal and information processing from the Beijing University of Technology, Beijing, China, in 2005 and 2008, respectively, and the Ph.D. degree in information technologies from University of Sydney, NSW, Australia, in 2014. He was a visiting scholar with the Biomedical and Multimedia Information Technology Research Group, School of Information Technologies, University of Sydney, NSW, Australia. He joined the Lenovo R$\&$T as a researcher of Image and Visual Computing Laboratory. He is now a researcher of Tencent, conducting pathological image processing for AI healthcare. His current research interests include video and image processing, pattern recognition, machine learning, computer vision, and deep learning.
\end{IEEEbiography}

\begin{IEEEbiography}[{\includegraphics[width=1in,height=1.25in,clip,keepaspectratio]{xiayong-eps-converted-to}}]{Yong Xia}
Yong Xia (S’05-M’08) received the B.E., M.E., and Ph.D. degrees in computer science and technology from Northwestern Polytechnical Uni-versity, Xi’an, China, in 2001, 2004, and 2007, respectively. He was a Postdoctoral Research Fellow in the Biomedical and Multimedia In-formation Technology Research Group, School of Information Technologies, University of Sydney, Sydney, Australia. He is currently working as a full professor at the School of Computer Science, Northwestern Polytechinical University. His research interests include medical imaging, image processing, computer-aided diagnosis, pattern recognition, and machine learning.
\end{IEEEbiography}

\begin{IEEEbiography}[{\includegraphics[width=1in,height=1.25in,clip,keepaspectratio]{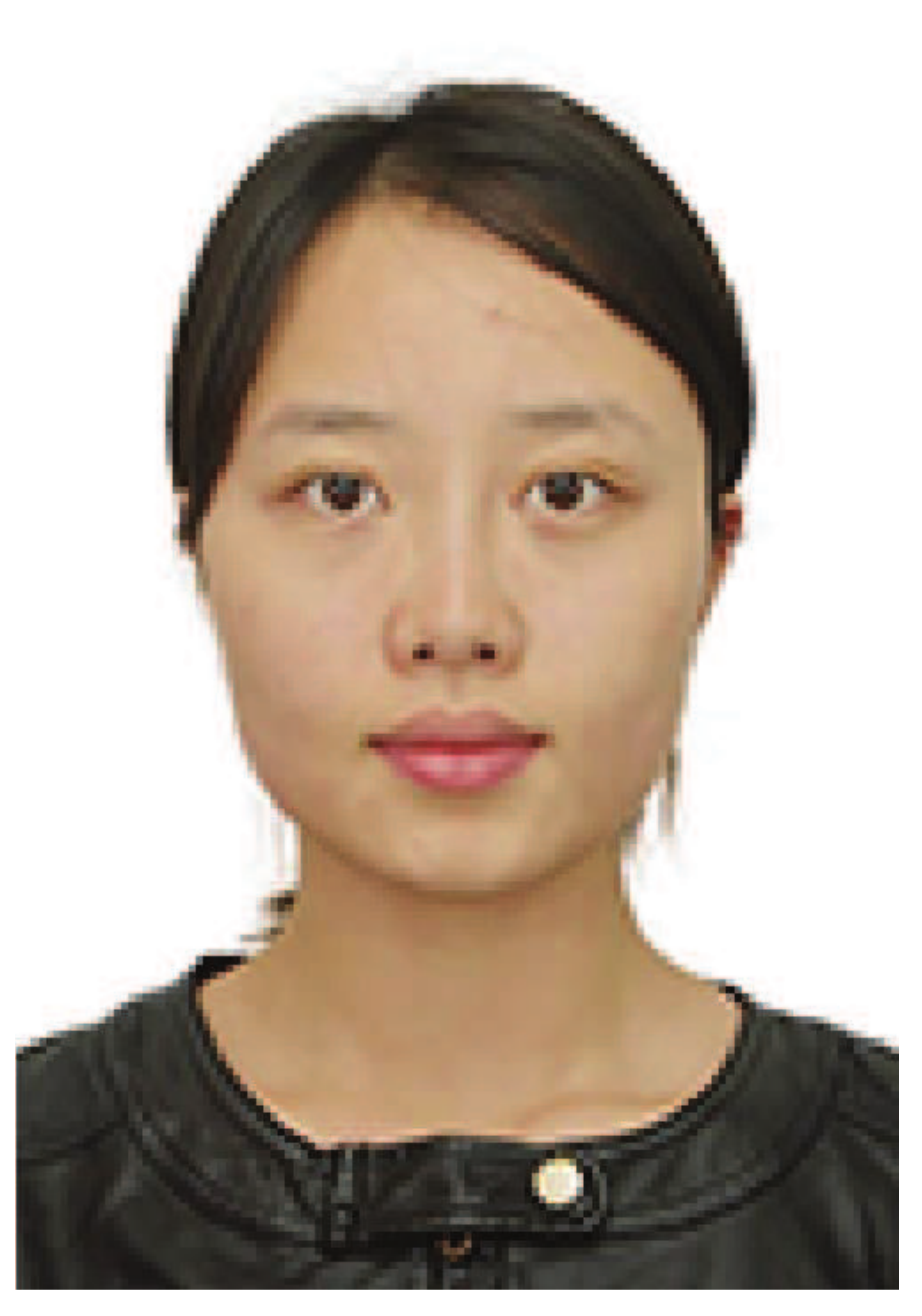}}]{shanshan wang}
	Shanshan Wang Shanshan Wang received her dual PhD degree in information technologies and biomedical engineering from the University of Sydney and Shanghai Jiao Tong University. She is an associate professor in Paul C. Lauterbur Research Center for Biomedical Imaging, Shenzhen Institutes of Advanced Technology, Chinese Academy of Sciences. Her research interests include machine learning, fast medical imaging and radiomics. She has published over 40 journal and conference papers in these areas. She is a member of the IEEE.
\end{IEEEbiography}

\end{document}